\documentclass[aps,pre,groupedaddress,twocolumn]{revtex4-2}
\usepackage[utf8]{inputenc}

\usepackage{graphicx}
\usepackage{amsmath}

\usepackage{multirow}

\newcommand{\D}{\mathrm{d}}

\usepackage{xcolor}

\usepackage{soul}

\begin{document}

%\preprint{}

\title{Coalescent dynamics of planktonic communities}


\author{Paula Villa Mart{\'\i}n}
\thanks{These authors contributed equally to this work.}
\author{Anzhelika Koldaeva}
\thanks{These authors contributed equally to this work.}
\author{Simone Pigolotti}
\email[]{simone.pigolotti@oist.jp}
%\thanks{}
\affiliation{Okinawa Institute of Science and Technology Graduate University, Onna, Okinawa 904-0495, Japan}


\date{\today}

\begin{abstract}
Planktonic communities are extremely diverse and include a vast number of rare species. The dynamics of these rare species is best described by individual-based models.  However, individual-based approaches to planktonic diversity face substantial difficulties, due to the large number of individuals required to make realistic predictions. In this paper, we study diversity of planktonic communities by means of a spatial coalescence model, that incorporates transport by oceanic currents. As a main advantage, our approach requires simulating a number of individuals equal to the size of the sample one is interested in, rather than the size of the entire community. By theoretical analysis and simulations, we explore the conditions upon which our coalescence model is equivalent to individual-based dynamics. As an application, we use our model to predict the impact of chaotic advection by oceanic currents on biodiversity. We conclude that the coalescent approach permits to simulate marine microbial communities much more efficiently than with individual-based models.
\end{abstract}

%\keywords{}

\maketitle

\section{Introduction}\label{sec:introduction}

Ecological communities are made up of a large number of species. Their diversity varies in space and time as a result of the ecological forces they are subject to \cite{levin1992problem}. In diverse ecological communities, one typically encounters a few very abundant species and many rare species, some of which are represented by just a few individuals. Because of these rare species, the diversity of ecological communities is best described by spatially explicit individual-based models (IBMs) \cite{cencini2012ecological,pigolotti2018stochastic}, rather than models based on species concentration or densities. IBMs have contributed to rationalize fundamental biodiversity patterns in terrestrial ecosystems, such as the scaling of the average number of encountered species with the size of the sampled area \cite{durrett1996spatial,rosindell2007species,pigolotti2009speciation,cencini2012ecological,pigolotti2018stochastic}. 

In prototypical spatially explicit IBMs such as the multi-species voter model \cite{durrett1994stochastic,durrett1996spatial,pigolotti2009speciation,pigolotti2018stochastic}, individuals are placed on a two-dimensional lattice and stochastically reproduce, die, and disperse.  Even simple spatial IBMs are challenging to solve analytically \cite{pigolotti2018stochastic}. A significant advancement in their understanding originates in the concept of {\em duality} \cite{durrett1994stochastic,durrett1996spatial}. In this context, ``duality'' is a mapping between the IBM and a different model, whose dynamics proceeds backward in time. For this reason, we often refer to the original model as the ``forward'' model and its dual as the ``backward'' model. The backward model considers a sample of the population of the forward model at a very long time and seeks to reconstruct its species composition. Individuals in the sample are represented as particles that evolve backward in time. If two particles happen to be on the same site, they can coalesce, signaling that the two corresponding individuals have a common ancestor and are therefore conspecific. Due to these events, the backward model is also termed ``coalescence'' model. By tracking coalescence events, the backward dynamics reconstructs the species composition of the original sample. 

In short, duality maps a spatial IBM into a system of coalescing random walkers. The advantage of this mapping is twofold. First, mathematical results have been obtained for systems of coalescing random walkers \cite{cox1989coalescing,bramson1991asymptotic}, leading to exact predictions of  biodiversity patterns \cite{pigolotti2018stochastic}. Second, backward models are much more efficient than forward models to simulate on a computer  \cite{rosindell2007species,pigolotti2009speciation,cencini2012ecological,pigolotti2018stochastic}. One reason is that backward simulations describe samples of a virtually infinite population, so that one does not have to worry about finite-size effects.  

IBMs have also been used to study microbial planktonic communities \cite{toroczkai1998advection,karolyi2000chaotic,hernandez2004clustering,pigolotti2012population,pigolotti2013growth}.  These models have been used to predict, for example, how fluid flows affect the fate of mutants characterized by a reproductive advantage \cite{pigolotti2012population,pigolotti2013growth,herrerias2018stirring,plummer2019fixation,guccione2021strong} or by a different diffusivity \cite{heinsalu2013clustering,pigolotti2014selective,pigolotti2016competition,singha2020fixation}. However, when used to predict biodiversity patterns, these models face severe computational limitations. Even state-of-the-art approaches are usually limited to communities of tens of thousands of individuals. For comparison, a liter of oceanic water can contain tens or hundreds of millions of planktonic cells \cite{bainbridge1957size}. To encompass this problem, it has been suggested that each individual in a IBM can be considered as a representative of an entire subpopulation \cite{scheffer1995super}. It is however unclear whether this interpretation can account for the dynamics of very rare species.

With this motivation in mind, we recently proposed a coalescence model for the dynamics of microbial planktonic communities \cite{villa2020ocean}, which encompasses the limitation of IBMs. In this paper, we study the dynamics of this coalescence model and argue that, under certain conditions, the coalescence model is dual to an IBM. We support these mathematical predictions with extensive numerical simulations of both models. We finally present an application our modeling approach to observational data from protist communities \cite{villa2020ocean}.

The paper is organized as follows. In Section~\ref{sec:model}, we introduce the (forward) IBM and the (backward) coalescence model. We prove that, in the weak noise limit, these two models are dual. In Section~\ref{sec:diversity}, we briefly introduce the main observables that are commonly employed in ecology to quantify biodiversity. In Section~\ref{sec:results}, we test our theory by extensive numerical simulations of the forward and backward models, both in the presence and absence of chaotic advection. We also compare the model prediction with metabarcoding data. Section~\ref{sec:conclusions} is devoted to conclusions and perspectives.

\section{Models}\label{sec:model}

In this section we introduce two approaches for modeling the diversity  of planktonic populations. 

The first approach  is via an IBM, in which an initial population stochastically evolves forward in time as a result of reproduction events, competition among individuals, advection--diffusion of individuals in space, and mutation/speciation events that give rise to new species. The forward model can be seen as the multi-species version of a two-species competition model \cite{pigolotti2012population,pigolotti2013growth} that does not include mutations. The second approach is based on a backward (coalescence) model \cite{villa2020ocean}. The backward model considers a sample of $N_s$ individuals and seeks to reconstruct its species composition by tracing the ancestry of the individuals backwards in time. 

We conclude the section by defining two regimes (weak and strong noise) characterizing the forward model. We then demonstrate that duality between the forward and the backward models rigorously holds in the weak noise regime.

\subsection{Forward model} 

A microbial population inhabits a two-dimensional square area $A=L \times L$, representing an aquatic environment.  Initially, individuals are homogeneously distributed. They can belong to different species, but for simplicity we neglect their species identity for the time being. Individuals can stochastically die, reproduce, and displace. As a result of these events, the total number of individuals $N(t)$ fluctuates in time. Each individual asexually reproduces at rate $\lambda$. When a reproduction occurs, the daughter individual is placed at a random position in a square of side $l$ centered on the mother position. From now on, we refer to this square as the ``neighborhood'' of an individual. Individuals die in a density-dependent way with rate $\lambda\,\hat{n}$, where $\hat{n}$ is the number of other individuals in their neighborhood. The dependence of the death rate on the local density represents competition. 

Each individual moves in space according to the advection-diffusion equations
\begin{eqnarray}\label{eq:forward}
\frac{\D}{\D t}x&=&v_x(x,y,t) +\sqrt{2D}\xi_x(t), \nonumber\\
\frac{\D}{\D t}y&=&v_y(x,y,t) +\sqrt{2D}\xi_y(t)\, ,
\end{eqnarray}
\vspace{6mm}
\begin{figure}[htb]
    \includegraphics[width=0.5\textwidth]{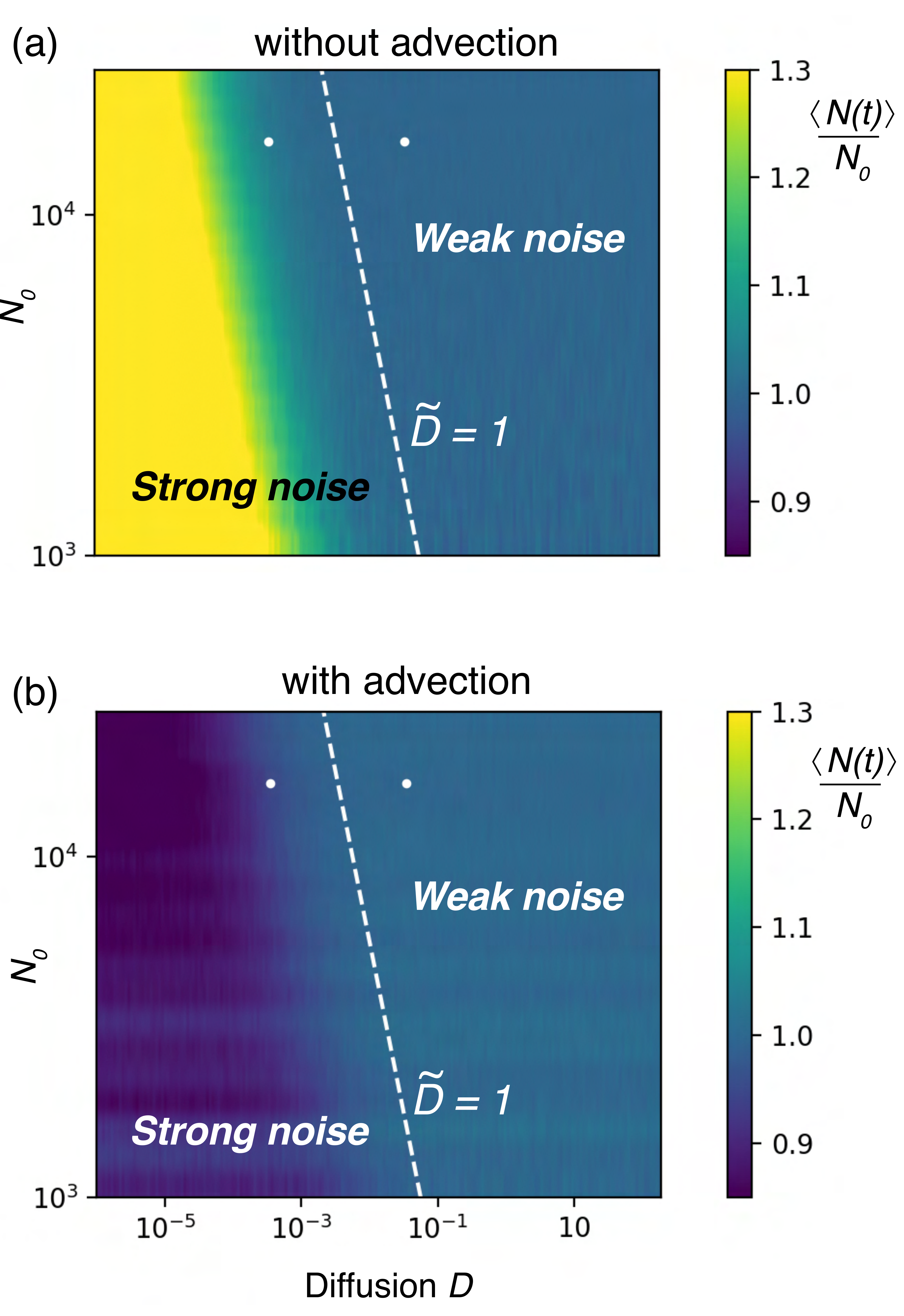}
    \caption{Normalized average number of individuals $\langle N(t)\rangle/N_0$ in (a) absence and (b) presence of advection. 
    The stationary population size $N_0$ is varied by tuning the neighborhood linear size $l$. The dashed line marks the theoretical condition $\tilde{D}=1$ separating the weak noise regime from the strong noise regime. In all simulations, we fixed $L=\lambda=1$.
    In both panels, two dots at $\tilde{D}=0.1$ and $\tilde{D}=10$ (with $N_0=16384$) mark the set of parameters chosen for our further analysis.}
  \label{fig:forward_model_snapshots}
\end{figure}
where $x$ and $y$ are the coordinates of the given individual. The terms proportional to $\sqrt{2D}$ represent diffusion. The functions $\xi_x(t), \xi_y(t)$ are white noise sources satisfying $\langle\xi_i(t)\rangle=0$, $\langle \xi_i(t)\xi_j(t')\rangle=\delta_{ij}\delta(t-t')$ where $i\in\{x,y\}$ and $\langle\dots\rangle$ denotes an average over realizations. For the time being, we impose periodic boundary conditions. The functions $v_x(x,y,t)$ and $v_y(x,y,t)$ represent an advecting fluid flow. In this paper we only consider incompressible velocity fields, $\vec{\nabla}\cdot \vec{v}=0$ where $\vec{v}=(v_x,v_y)$ and $\vec{\nabla}=(\partial/\partial x, \partial/\partial y)$. Indeed, two-dimensional velocity fields in the oceans are incompressible to a very good approximation, although in some local regions vertical currents can alter this picture \cite{thomas2008submesoscale}, potentially impacting population dynamics \cite{pigolotti2012population,benzi2012population,plummer2022oceanic}.

In the absence of birth and death dynamics and thanks to incompressibility, Eqs.~\eqref{eq:forward} predicts a homogeneous stationary distribution of individuals. We tentatively assume that this distribution remains homogeneous in the presence of birth and death processes. We call $N_0$ the average population size under this hypothesis. By imposing that birth events statistically balance death events, we find that 
\begin{equation}\label{eq:n0cond}
N_0=L^2/l^2,
\end{equation} 
i.e. the average population size is equal to the ratio between the system area and the area of the interaction neighborhood, see Appendix A. In practice, this means that each interaction neighborhood contains one individual, on average. In the following, we take this value as the initial population size, $N(0)=N_0$. 

We want to understand whether the assumption of homogeneous density holds. To this aim, we study a macroscopic description of our IBM, see Appendix B. We find that the populations remain homogeneous in the case when the stochastic fluctuations induced by birth and death processes are relatively small. In this case, the average number of individuals does not significantly deviate from $N_0$.

In two dimensions, the relative strength of fluctuations is controlled by the dimensionless parameter
\begin{equation}\label{eq:tilded}
\tilde{D}=D N_0/(\lambda L^2),
\end{equation}
see \cite{pigolotti2013growth}. For $\tilde{D}\gg 1$, stochastic fluctuations are small. In contrast, for $\tilde{D}\ll 1$, fluctuations dominate the dynamics.  In the following, we refer to the $\tilde{D}>1$ and  $\tilde{D}< 1$ cases as the ``weak noise'' and ``strong noise''regime, respectively. 
. 

Simulations of the model confirm that, in the strong noise regime, the average number of individuals significantly differ from $N_0$, see Fig.~\ref{fig:forward_model_snapshots}. This means that our assumption that the birth-death dynamics does not affect the average number of individuals breaks down. 

In particular, in the absence of advection, our simulations show that the average number of individuals exceeds $N_0$, see Fig.~\ref{fig:forward_model_snapshots}a. This result contrasts with that in \cite{pigolotti2013growth}, where a reduction of the average number of individuals was observed in the strong noise regime. This discrepancy can be explained by a difference in the way birth is implemented in the two models: here, the daughter cell is placed at a random position in the neighborhood of her mother, whereas in \cite{pigolotti2013growth} the daughter cell is placed at the same position as the mother. 

In contrast, in the presence of advection, the average population size decreases in the strong noise regime, see Fig.~\ref{fig:forward_model_snapshots}b.  A qualitative explanation is that advection effectively precludes individuals to visit some regions, thereby increasing effective competition.

Hereafter, we fix $\lambda=1$, $L=7.5$, and $N_0=16384$. The value of $l$ corresponding to these choices is obtained from Eq.~\eqref{eq:n0cond}.

\begin{figure}[htb]
    \includegraphics[width=0.45\textwidth]{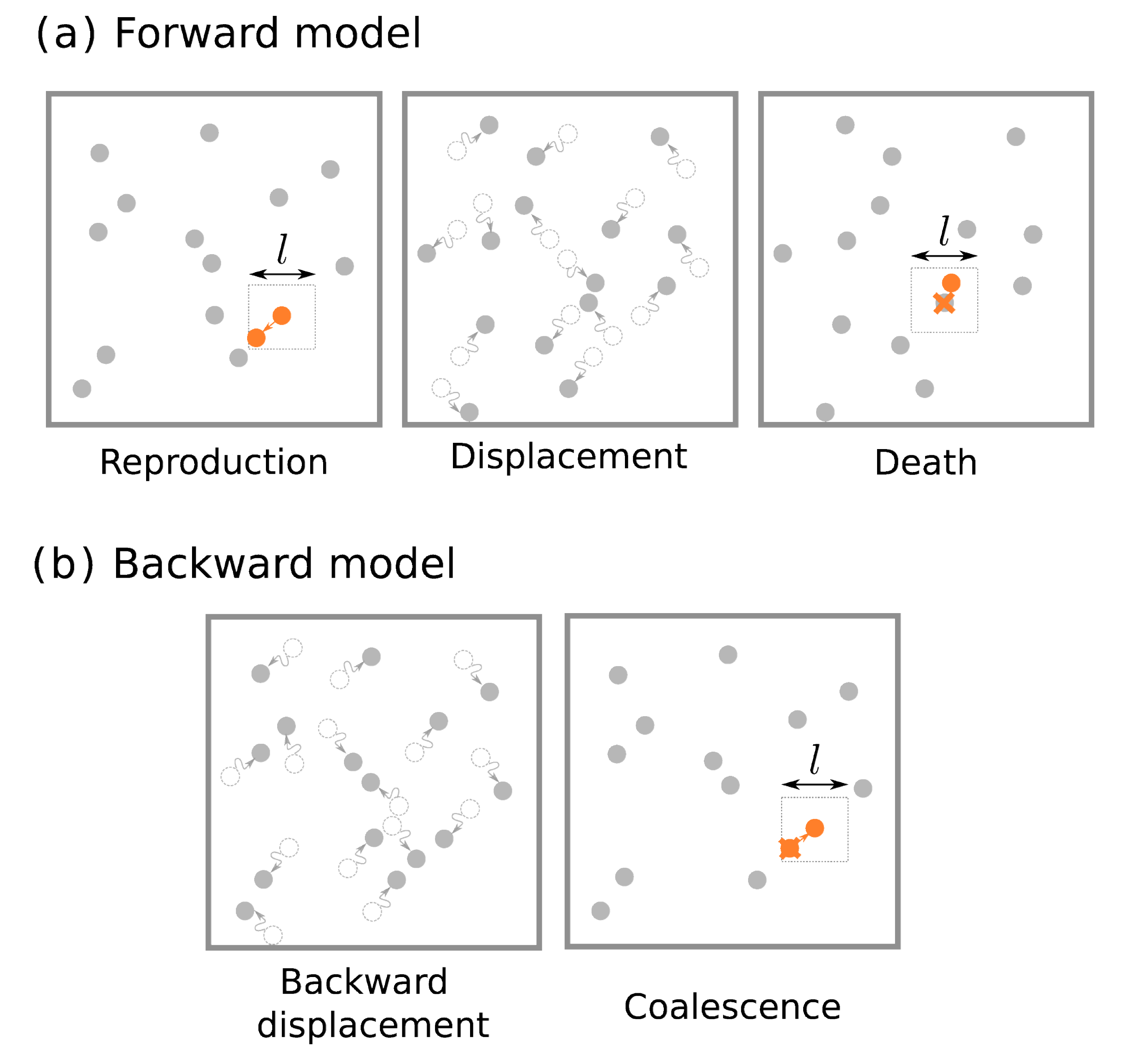}
    \caption{{\bf{Events in the forward and backward models}}. (a) In the forward model, individuals reproduce at rate $\lambda$ within a neighborhood, displace, and die with rate $\lambda\,\hat{n}$, where $\hat{n}$ is the number of other individuals in their square neighborhood of linear size $l$. (b) In the backwards model individuals displace and coalesce with rate $\bar{\lambda}$ with individuals within their neighborhood. In the weak noise regime, the two models are dual under the condition $\lambda=\bar{\lambda}$.
    }
  \label{fig:forward_model}
\end{figure}

\subsection{Backward model and duality}
\label{sec:duality}

The backward model (or coalescence model) describes the dynamics of $N_s$ Lagrangian tracers. These tracers represent a sample of individuals from a final population, i.e. from a population that evolved according to the forward model at a time $t_f\gg0$. We assume that the sample is homogeneously distributed in a sample area  $L_s\times L_s$, with $L_s\le L$. We trace back in time the evolution of the tracers in the sample.  The coordinates of the tracers evolve according to Eq.~\eqref{eq:forward}, which we integrate with negative time increments. 

While evolving back in time the coordinate of a given tracer, we might reach the time at which the individual represented by the tracer was born. From that instant, the tracer represents the position of the mother of the chosen individual. This means that, at a given time, each tracer represents either an individual in the final population or one of its ancestors.  If the mother of the chosen individual (or one of her ancestors) is alive in the final population, this implies that the two tracers must be in the neighborhood of each other at the time in which the birth event occurs. If this is the case, we say that a ``coalescence'' has occurred, and the two tracers are merged into one. 
%An example of evolving trajectories simulated with the backward model in the absence and presence of the velocity field is shown in Fig.~\ref{fig:trees}.

The events occurring in the forward and backward models are summarized in Fig.~\ref{fig:forward_model}a and Fig.~\ref{fig:forward_model}b, respectively.

To fully specify the coalescence model, we need to determine the rate $\bar{\lambda}$ at which two individuals coalesce if they are in the neighborhood of each other. We do so by considering a situation in which $N_s=N_0$, i.e., the number of tracers is equal to the average number of individuals in the forward model. In this scenario we are tracing all individuals, therefore the total rate of birth $\lambda N_0$ in the forward model must match the total rate of coalescence in the backward model. This latter rate is equal to $\bar{\lambda}N_s$ by the same argument made in Appendix A. Imposing that the two total rates must be equal and using $N_s=N_0$ leads to $\bar{\lambda}=\lambda$, i.e. the coalescence rate for individuals in the same neighborhood should match their birth rate in the forward model. 

We remark that this argument relies on the assumption that the dynamics of the forward model is in the weak noise regime. In the strong-noise regime, some of the assumptions underlying duality do not hold. First, we can not assume that a sample of individuals in the final population is homogeneously distributed. Second, since in the strong-noise regime $\langle N(t)\rangle\neq N_0$, we can not easily draw a correspondence between the rates of the forward model and the backward model.

\subsection{Mutations and multi-species dynamics}

We now add to the forward and backward models the notion of species identity. In the forward model, we assume that a newborn individual belongs to a new species with probability $\mu$. In ecological terms, the probability $\mu$ can either be interpreted as a speciation probability, or as a probability for individuals to be replaced by individuals belonging to new species immigrating from outside the community. If the model is used to interpret metabarcoding studies, $\mu$ represents the probability for individuals to accumulate sufficient mutations to be considered as a new operational taxonomic unit (OTU).

\vspace{6mm}
\begin{figure}[htb]
    \includegraphics[width=0.45\textwidth]{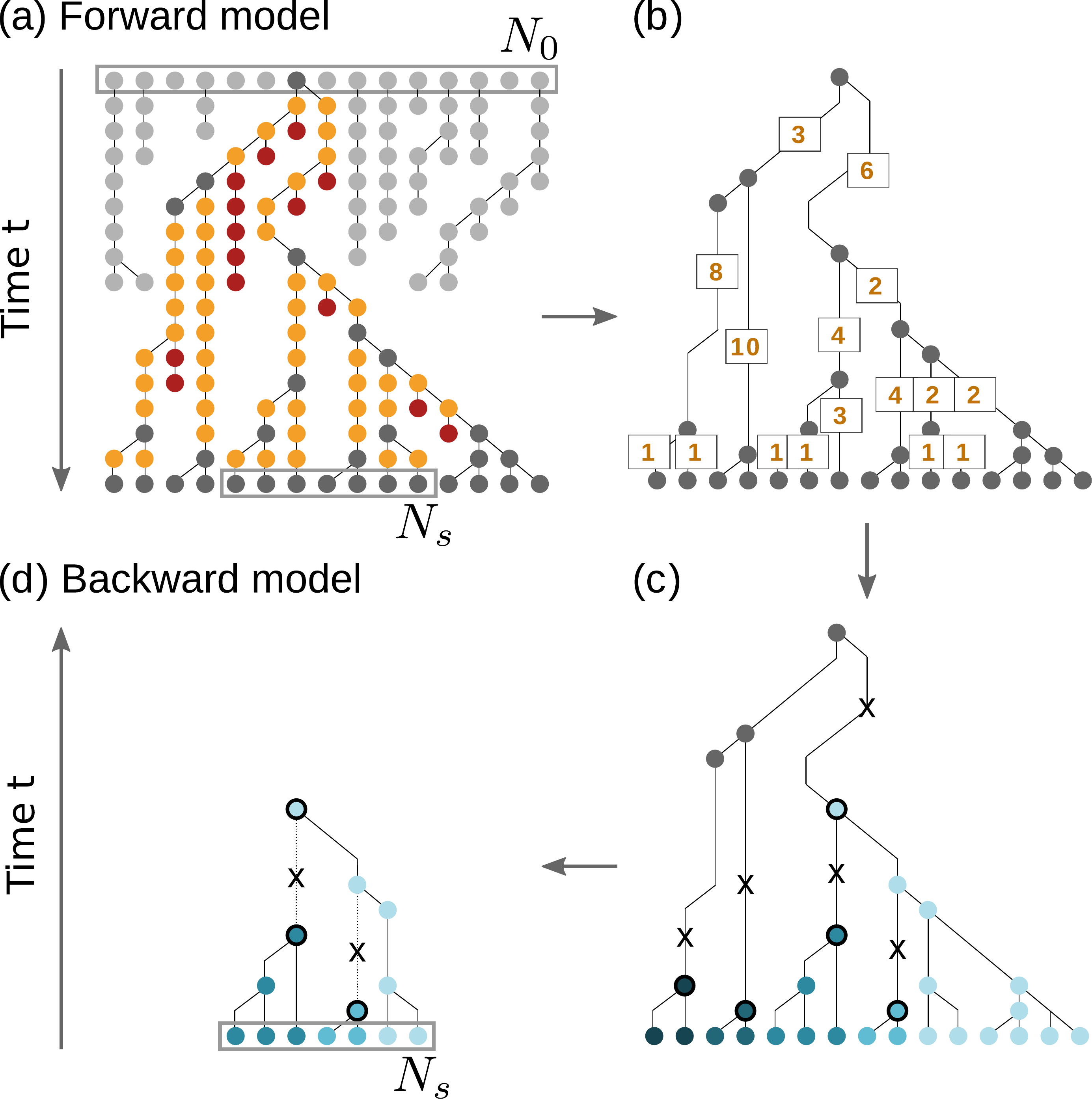}
    \caption{{\bf{Ancestry trees}}. (a) Dynamics of the forward model represented as a tree. An initial set of $N_0$ individuals reproduce and die over time until all remaining individuals share the same common ancestor. Individuals that die before the final time (dark red dots)
    are not relevant for the final population, and are therefore removed. Individuals that do not have descendant in the final population (light gray dots) are removed as well. (b) individuals that have one direct descendant in the remaining tree (orange dots) are removed; their number are saved as variables $d_i$  for each branch $i$. (c) Mutations (marked with Xs) are introduced at each tree branch $i$ with a probability $p_i=1-(1-\mu)^{d_i}$, see \cite{pigolotti2018stochastic}.
    (d) Ancestry tree of the backward model. In this case, we consider a sample of $N_s$ individuals from the final population and reconstruct its ancestry. Individuals can coalesce or mutate until one individual remains. This approach directly ignores branches corresponding to individuals that do not belong to the sample.}
  \label{fig:ancestry_tree}
\end{figure}

When simulating the forward model, one can keep track of species identity by assigning each newborn individual to a new species with probability $\mu$. We however adopt an equivalent, but more efficient strategy, see \cite{pigolotti2018stochastic,rosindell2008coalescence}.  We evolve individuals without keeping track of their species identity, until all individuals descend from a single individual in the original population, see Fig.~\ref{fig:ancestry_tree}a. The collection of descendents of this individual constitutes the ancestry tree. Individuals that do not belong to the ancestry tree (light grey in Fig.~\ref{fig:ancestry_tree}a) do not affect the diversity of the final population and can therefore be ignored. A further simplification is to remove individuals that have only one descendant in the ancestry tree (yellow in Fig.~\ref{fig:ancestry_tree}a). This can be done by keeping track of the number of duplications $d_i$ occurred in each branch $i$ separating the remaining individuals,  see Fig.~\ref{fig:ancestry_tree}b. Since at each duplication event the probability of a mutation is equal to $\mu$, the total probability that at least a mutation has occurred in a branch $i$ is equal to $p_i=1-(1-\mu)^{d_i}$. In this way, we  assign the mutations {\em a posteriori}, by drawing from the probabilities that a mutation has occurred in each branch, see Fig.~\ref{fig:ancestry_tree}c. 

In the backward model, mutations occur continuously at a stochastic rate $\mu \lambda dt$. The reason is that in the backward model we do not track individual birth events, unless they lead to coalescence. As anticipated, the advantage of the coalescence model is that it is possible to reconstruct the identity of a sample of $N_s$ individuals in the final population being a subset of the total population. The corresponding ancestry tree is a subset of the corresponding tree in the forward model which includes all ancestors of the individuals in the sample up to their most recent common ancestor, see Fig.~\ref{fig:ancestry_tree}d. In practice, mutations can be assigned on the branches of this tree in a similar way as for the forward model. However, this is not necessarily an efficient procedure in this case. The reason is that, for large system size and in the presence of advection, the time it takes for all tracers to coalesce can be exceedingly long. At the same time, it is not necessary to track tracers backward in time before a mutation, since events occurred before the mutations do not affect the diversity of the final sample. For this reason, we evaluate the probability of mutations at each time step of the backward dynamics and we simply eliminate tracers that have mutated. 

We compare predictions of both models for a sample of $N_s$ individuals taken at a final large time $t_f$. We fix $N_s<N_0$. In the absence of advection, since we might have $\langle N(t)\rangle \gtrsim N_0$, we fix $N_s=16000$. In the presence of advection, population sizes are $\langle N(t)\rangle \lesssim N_0$, and we consider $N_s=8192$ to ensure $N_s<N_0$.

\section{Diversity measures}\label{sec:diversity}

\begin{figure*}[htb]
    \includegraphics[width=0.99\textwidth]{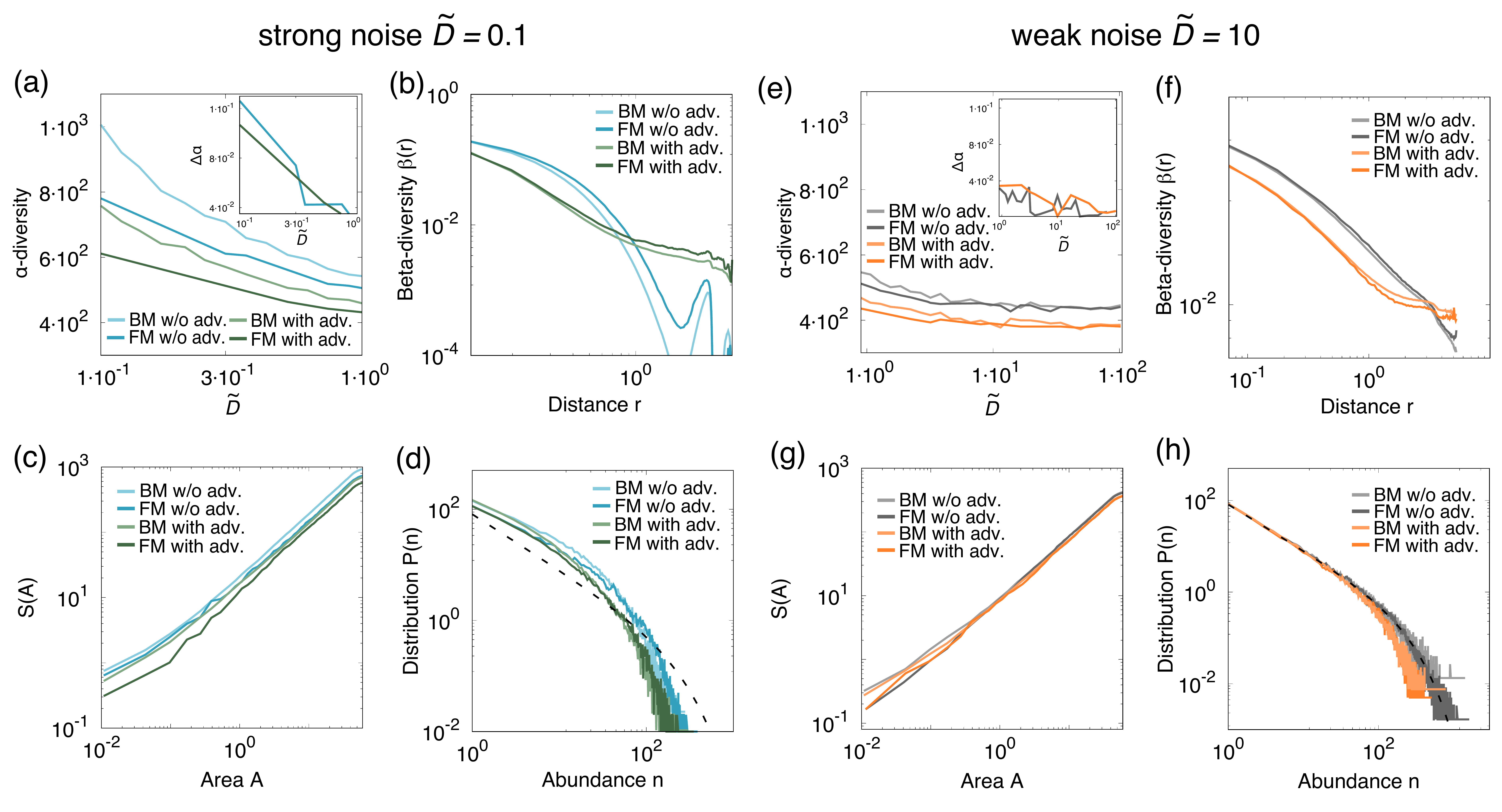}
    \caption{{\bf{Diversity measures predicted by the models with and without advection in the strong (a, b, c, d)  and weak (e, f, g, h) noise regimes.}} (a, e) Value of
    $\alpha$-diversity as a function of the parameter $\tilde{D}$. Measures for forward model (FM) (darker colors) and backward model (BM) (lighter colors)   are shown. (b, f) $\beta-$diversity, (c, g) Species-Area $S(A)$, and (d, h) species abundance distribution for $\tilde{D}=0.1,10$. In these panels, dark/light colors show curves for the forward/backward model for $\tilde{D}=10$; these curves nearly coincide.  Dashed lines of panels (d) and (h) show the analytical prediction for a well-mixed system, Eq.~\eqref{eq:fisher}. In all panels, the mutation probability is equal to $\mu=5 \cdot 10^{-3}$.}
  \label{fig:div_meas}
\end{figure*}

Quantifying biodiversity by simply counting the number of competing species is not always appropriate. The reason is that most ecological communities, including planktonic ones, are composed of relatively few abundant species and a large number of rare species. The definition of biodiversity in terms of number of species is insensitive to this distinction. This definition also relies on the possibility of observing all these rare species, which is often impossible in practice. To address this issue, other measures of biodiversity have been proposed in the literature \cite{xu2020diversity}. These alternative definitions give different weight to abundant and rare species.

As anticipated, a basic biodiversity measure is the number of species present in the community. Other measures take into account more explicitly spatial distributions and compositional heterogeneity. Specifically, at equal number of species, a population can be equally distributed among species or be dominated by a few of them. Moreover, the manner that species distribute in space, i.e. whether individuals of a species are grouped within specific areas or highly dispersed in space, is an important characterization of biodiversity.  Here, we review the most common measures of biodiversity:

\begin{itemize}
\item\textbf{$\alpha$-diversity}. The $\alpha$-diversity is the first and simplest measure. It is defined as the total average number of species in the community \citep{whittaker1960vegetation}. We measure the $\alpha$-diversity by simply averaging the total number of species over multiple realizations of the forward and backward models.

\item\textbf{$\beta$-diversity}. The $\beta$-diversity quantifies spatial correlations within species. It describes how species composition changes from local to larger scales. There exist slightly different definitions of $\beta$-diversity in the literature \cite{tuomisto2010diversity1,tuomisto2010diversity2}. We 
define $\beta$-diversity as the probability that two random individuals at a given distance $r=\sqrt{x^2+y^2}$ belong to the same species 
\begin{equation}
\beta(r)=\displaystyle\frac{\sum_i N_{s_i,s_i}(r)}{\sum_{i,j} N_{s_i,s_j}(r)}
\end{equation}
where $N_{s_i,s_j}(r)$ is the number of pairs $i,j$ of species $s_i,s_j$ at distance $r$ \cite{pigolotti2018stochastic}.
At equal number of species, highly grouped communities are characterized by a steeper $\beta$-diversity
than more dispersed ones.

\item\textbf{Species-area relation}. The species-area relation is defined as the average number of species $S$ 
found in an area $A$ \citep{rosenzweig1995species}. The species-area relation has been fitted by mathematical relations of the form $S(A)\propto A$ for small and large scales, and $S(A)\propto A^{z}$ for intermediate scales \cite{preston1960time,rosenzweig1995species,hubbell2001unified,arrhenius1921species}. 
We average the final number of species $S$ over several realizations of our models for increasing sampling areas of size $A=L_s\times L_s$ located
at the center of our system.

\item\textbf{Species-abundance distribution}. The species-abundance distribution (SAD) quantifies the the compositional heterogeneity of a population in terms of the relative abundance of species. It is defined as the frequency $P(n)$ of species with abundance $n$ in a sample. In a well-mixed system of population size $N_0$ and with mutation probability $\mu$, the species-abundance distribution has the form
\begin{equation}\label{eq:fisher}
P(n)=\displaystyle\frac{N_0\mu e^{-\mu n}}{n},
\end{equation}
see \cite{hubbell2001unified,volkov2003neutral}. We expect our spatially-explicit model to generate a similar species-abundance distributions, at least in the high diffusion limit. Apart from this limiting case, analytical predictions for the species-abundance distribution in spatially explicit models are rather hard to obtain \cite{pigolotti2018stochastic}. The species-abundance distribution is normalized to the $\alpha$-diversity, i.e., the average total number of species in the community:
\begin{equation}\label{eq:alpha_mf}
S\approx \int_1^\infty dn \, \frac{N_0\mu e^{-\mu n}}{n}.
\end{equation}
\end{itemize}

\section{Results}\label{sec:results}

In this section, we numerically verify that the forward and backward models are equivalent in the weak noise regime, and test whether this equivalence extends in the strong noise regime. We also study whether this equivalence is affected by the value of the mutation probability $\mu$. To these aims, we extensively simulate both systems for a broad parameter range and compute diversity measures of the final population averaged over $10^3$ realizations. We perform these comparisons both in the presence and in the absence of advection.

\subsection{Biodiversity measures in the absence of advection}\label{sec:effectdiffusion}

We consider the two models in the weak and strong noise regimes and first focus on the $\alpha$-diversity for different diffusion rates $D$.  As expected, the predictions of the forward and backward models present a small by significative discrepancy in the strong noise regime, see Fig.~\ref{fig:div_meas}a. We quantify this discrepancy by the absolute relative difference 
\begin{equation}\label{eq:deltaalpha}
\Delta \alpha=\frac{|\alpha_f-\alpha_b|}{\alpha_f+\alpha_b},
\end{equation}
where $\alpha_f, \alpha_b$ are the $\alpha$-diversities measured in the forward and backward dynamics, respectively. In the weak noise regime, we observe compatible values of the $\alpha$-diversity, see Fig.~\ref{fig:div_meas}e. This analysis reveals that, for example for $\tilde{D}\approx 0.1$, the discrepancy is on the order of $10\%$, see inset of Fig.~\ref{fig:div_meas}a. In particular, the backward model predicts a higher number of species than the forward model. As discussed in section~\ref{sec:model} for the number of individuals, the differences between the forward and backward models in the strong noise limit can depend on model details, such as the microscopic implementation of mutations.

Comparisons of the $\beta$-diversity, Fig.~\ref{fig:div_meas}b,f, species-area relation, Fig.~\ref{fig:div_meas}c,g, and species-abundance distribution, Fig.~\ref{fig:div_meas}d,h lead to similar conclusions.  In the weak noise regime, the forward and backward models yield nearly identical predictions for all these quantities, as expected. In the strong-noise regime, we observed some discrepancies, of comparable magnitude as for the $\alpha$-diversity.

\subsection{Biodiversity measures in the presence of chaotic advection}\label{sec:effectadvection}

We now move to the case with advection. We define a two-dimensional incompressible advecting field in terms of a stream-function $\phi(x,y)$. The components of the field are related with the stream function by
\begin{align}
v_x(x,y;t)&=-\frac{\partial \phi(x,y;t)}{\partial y}\nonumber\\
v_y(x,y;t)&=\frac{\partial \phi(x,y;t)}{\partial x} .
\label{ad_field}
\end{align}
This definition automatically guarantees the incompressibility condition $\vec{\nabla}\cdot \vec{v}=0$. We choose a dimensionless stream function that generates a chaotic vortex in the vicinity of the meandering jet \cite{cencini1999mixing}:
\begin{equation}
\phi(x,y) = -\tanh \left( \frac{y - B(t)\cos(kx)}{\sqrt{1+k^2B^2(t)\sin^2(kx)}} \right) +cy,
\label{eq:stream}
\end{equation}
where $B(t) = B_0 + \epsilon \cos(wt+ \Phi)$. We fix
$k = 2 \pi/L$, $c = 0.12$, $L = 7.5$, $B_0 = 1.2$, $\epsilon = 0.3$, $w = 0.4$, $\Phi = \frac{\pi}{2}$. We impose periodic boundary conditions. We numerically solve the differential equations \eqref{eq:forward} with the advecting field \eqref{ad_field} using
a 4th order Runge-Kutta method.

As in the case without advection, we observe a significant discrepancy in the $\alpha$-diversity between  the  forward and backward models in the strong noise regime, see Fig.~\ref{fig:div_meas}a. The discrepancy progressively decreases for bigger $\tilde{D}$, see Fig.~\ref{fig:div_meas}e. The discrepancy tends to be smaller in the presence of advection rather than in the absence of it, see inset of Fig.\ref{fig:div_meas}a. One possible explanation for this difference is that, at equal diffusivity, the system subject to chaotic advection presents a different effective diffusivity \cite{biferale1995eddy}. Other diversity measures show the expected behavior, with a close correspondence in the weak noise regime and appreciable differences in the strong noise regime, see Fig.~\ref{fig:div_meas}b-d, and Fig.~\ref{fig:div_meas}f-h.

A comparison between Fig.~\ref{fig:div_meas}a and Fig.~\ref{fig:div_meas}e. shows that advection tends to reduce the average number of species. The choice of periodic boundary conditions crucially impacts this result. 

\subsection{Effect of mutation probability}

\begin{figure}[htb]
    \includegraphics[width=0.50\textwidth]{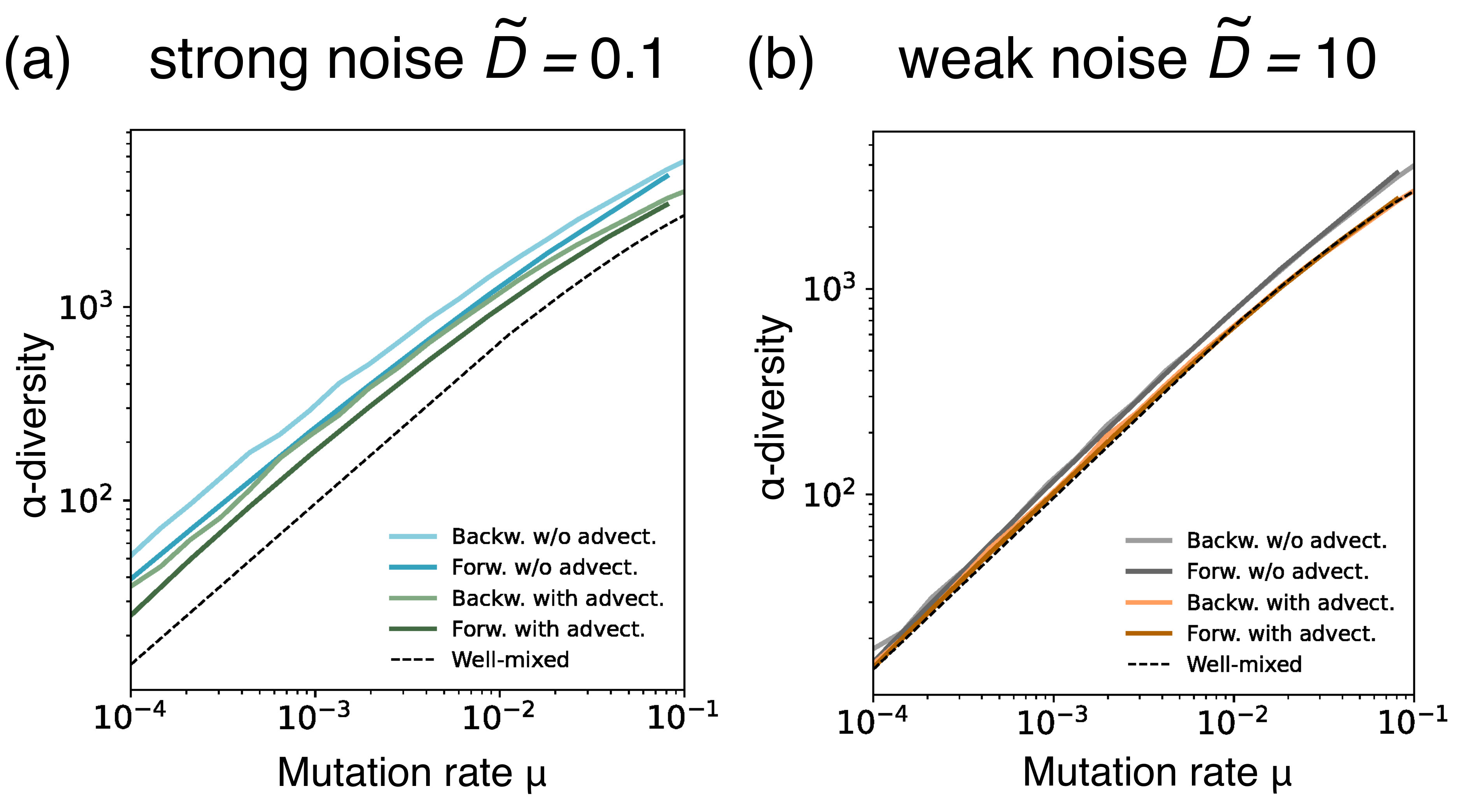}
    \caption{{\bf{$\alpha$-diversity as a function of the mutation probability}} in the strong  (a), and weak (b) noise regimes. Dashed lines show the analytical prediction for well-mixed systems, see Eq.~\eqref{eq:alpha_mf}.}
  \label{fig:alpha_mutation-rate}
\end{figure}

We now study the dependence of the $\alpha$-diversity on the mutation probabilities $\mu$ for both the forward and backward model. In the strong noise regime and for a low mutation probability, the forward model predicts a lower $\alpha$-diversity than the backward model, see Fig.~\ref{fig:alpha_mutation-rate}a and  Fig.~\ref{fig:alpha_mutation-rate}b.  The relative difference between forward and backwards models weakly depends on $\mu$.  In the weak noise regime, models are compatible for all values of the parameter $\mu$, see Fig.~\ref{fig:alpha_mutation-rate}a,b. These behaviors are qualitatively similar in the presence and in the absence of advection.

\begin{figure*}[htb]
    \includegraphics[width=0.99\textwidth]{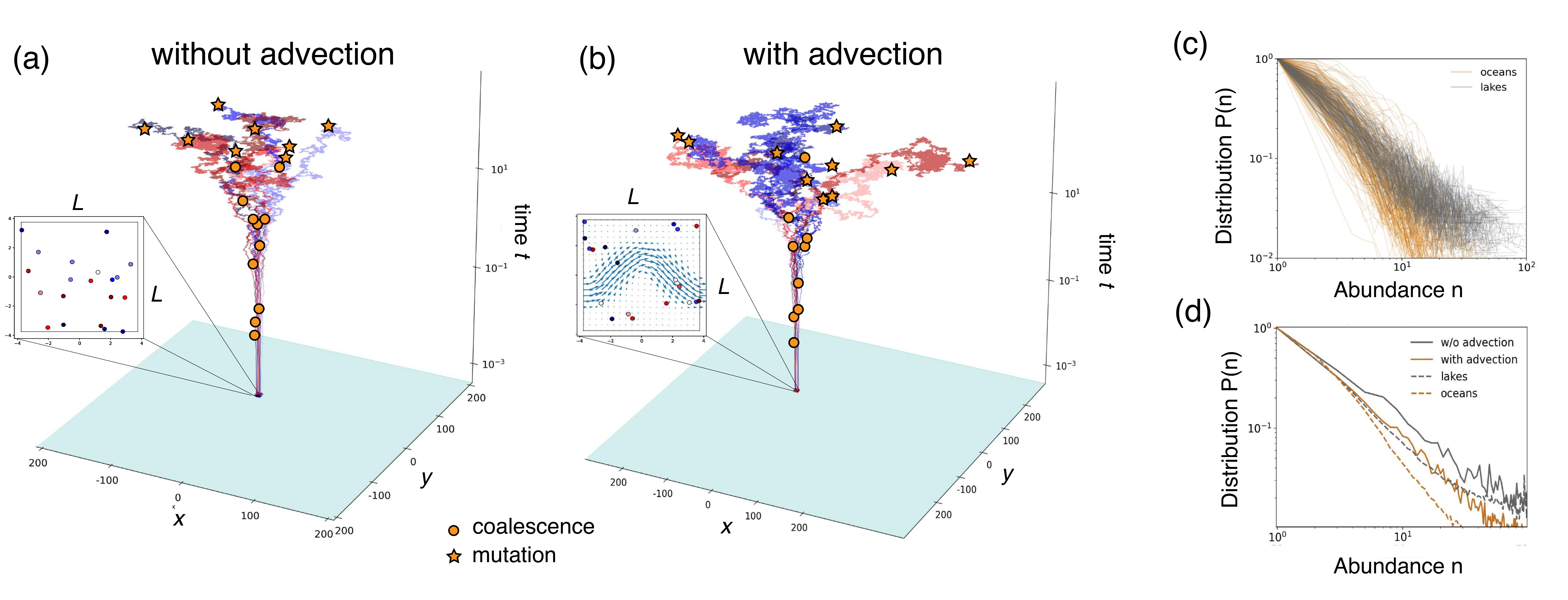}
    \caption{Individual trajectories modeled with the coalescence model in the (a) absence and (b) presence of advection. At the initial time $t=0$, the populations are homogeneously distributed in a square of size $L \times L$. We simulate the coalescence model with open boundary conditions. In (b), we employ the velocity field given in Eqs.~\eqref{ad_field} and \eqref{eq:stream}. Parameters are specified in Section~\ref{sec:results}B. (c) Species abundance distributions of individual protist populations sampled in oceans and lakes. (d)  Dashed curves represent average abundance distributions of protist populations sampled in oceans and lakes (c); solid curves correspond to numerical simulations of the backward models with open boundary conditions without and with advection, where we fixed $\tilde{D} = 1$. }
  \label{fig:trees}
\end{figure*}

\subsection{Metagenomic data}

In this section, we demonstrate that the coalescence model is able to predict species-abundance distributions observed in aquatic environments.

We numerically simulate the coalescence model in the presence and absence of advection, see Fig.~\ref{fig:trees}a, b. In this case, we adopt open boundary conditions as these are more relevant for the ocean, where a sample is embedded in a very large area. We do not perform a comparison with the forward dynamics, since the forward IBM can not be easily formulated with open boundaries. Individuals are homogeneously distributed in a square at $t = 0$. In the case with advection, we set $c = 0.12$, $B_0 = 1.2$, $w = 0.5$, and $\epsilon = 4$. 

We compare the model prediction for the SAD with observational data. To this aim, we employ two metabarcoding datasets: one from the TARA Oceans expedition \cite{ser2018ubiquitous}, and one from  freshwater lakes \cite{boenigk2018geographic}. Metabarcoding techniques permit to sample planktonic diversity  at unprecedented resolution \cite{de2015eukaryotic}. In metabarcoding studies, one obtains from a sample  DNA fragments corresponding to a highly conserved region of the genome, in this case a portion of the 18S ribosomal RNA gene. Since this region is highly conservffed, sequences in the sample with high degree of genetic similarity (at least 97\%, in this case) are likely to come from individuals within the same taxonomic group. These groups, identified by genetic similarity, are called operational taxonomic units (OTU).  We plot the abundance distributions of the observed OTUs from each dataset, see Fig.~\ref{fig:trees}c. The data demonstrates that oceanic currents promote abundance of rare species \cite{villa2020ocean}. 

Our coalescence model predicts that the effect of oceanic currents leads to SAD curves characterized by a steeper decay, see Fig.~\ref{fig:trees}d and \cite{villa2020ocean}. This effect is also observed in the data, see Fig.~\ref{fig:trees}d, although the slopes in the log-log plot are slightly different in the data. As extensively discussed in \cite{villa2020ocean}, the quantitative value of the slope is affected by microscopic details of the model, and aspects of the metabarcoding analysis such as the similarity threshold chosen to identify OTUs.

\section{Conclusions}\label{sec:conclusions}

In this paper, we developed a model for the dynamics of microbial aquatic communities based on the idea of coalescence. Our coalescence model predicts the diversity of a sample of organisms embedded in a very large, spatially extended populations. It  encompasses the limitations of individuals-based model in describing communities made up of huge number of individuals. Our model has the potential to bridges the gap between ecological dynamics at the individual level and large-scale spatial dynamics. 

We have shown that, in the weak noise regime, the model is equivalent to an individual-based model proceeding forward in time. In the strong-noise regime, this correspondence is only approximate, but both models predict qualitatively similar biodiversity patterns. Due to its advantages, the coalescence model presented in this paper provides a a versatile and powerful tool to predict biodiversity observed in metabarcoding studies of planktonic communities \cite{villa2020ocean}.

Although, for simplicity, we focus on a simple model of microbial competition dynamics, our approach can be extended to more general ecological settings and to other communities. Such generalizations,
combined with high-throughput sequencing data, have the potential to shed light on the main ecological forces determining community dynamics.

\begin{acknowledgements}
We thank Massimo Cencini for comments on a preliminary version of this manuscript. We thank Florian Pflug for helping us to debug the code. We thank Ales Bucek and Tom Bourguignon for the analysis of metabarcoding data and stimulating discussions.
We are grateful for the help and support provided by the Scientific Computing and Data Analysis section of Research Support Division at OIST.
\end{acknowledgements}

\appendix

\section{Stationary number of individuals in the weak noise limit}\label{app:n0}

In this Appendix, we compute the stationary number of individuals $N_0$ in the limit in which individuals are homogeneously distributed. The numer of individuals $n$ inside each neighborhood is a Poisson random variable with average $\mu=N_0/M$, where $M=(L/l)^2$ is the total number of neighborhoods.

To estimate $N_0$, we impose that the total average birth and death rates should balance. The total average birth rate is simply $\lambda N_0$. Individuals die with a rate proportional to the number of other individuals in each neighborhood. Therefore, the total average death rate is equal to
\begin{align}
\mbox{total average death rate}&=\lambda M \langle n(n-1)\rangle \nonumber\\
&=\lambda M \sum_n \frac{n(n-1) \mu^n e^{-\mu}}{n!}\nonumber\\
&=\lambda M\sum_n \frac{\mu^n e^{-\mu}}{(n-2)!}\nonumber\\
&=\lambda M\mu^2=\frac{\lambda N_0^2}{M}\nonumber\\
&=\frac{\lambda N_0^2 l^2}{L^2}.
\end{align}
Imposing that the total average death rate must be equal to the total average birth rate leads to the condition $L=l\sqrt{N_0}$, or equivalently $N_0=M$. 

\section{Macroscopic description of the forward model}\label{app:n00}

In this Appendix, we study a macroscopic description of our IBM. We introduce the density of individuals $n(x,y;t)$, defined so that its integral over a given area yields the number of individuals in that area at time $t$. We also define the concentration $c(x,y;t)=(L^2/N_0)n(x,y;t)$. The normalization factor $L^2/N_0$ ensures that the average concentration is equal to one, if the assumption of homogeneous density holds. 

The dynamics of the concentration $c(x,y;t)$ can be derived in the small noise limit, i.e. by assuming that the stochastic fluctuations induced by birth and death processes are relatively small. Under this assumption, the concentration is described by the stochastic Fisher-Kolmogorov equation
\begin{equation}\label{eq:fkpp}
\frac{\partial}{\partial t} c(x,y;t)=\lambda(c-c^2) -  \vec{\nabla}\cdot[\vec{v}c]+D\nabla^2 c+\sigma(c)\xi(x,y;t),
\end{equation}
where $\xi(x,y,t)$ is a noise field satisfying $\langle \xi(x,y,t)\rangle=0$ and $\langle \xi(x,y,t)\xi(x',y',t') \rangle=\delta(x-x')\delta(y-y')\delta(t-t')$. The multiplicative noise is interpreted using the Ito prescription; its amplitude is equal to $\sigma(c)=\sqrt{\lambda L^2 c(1+c)/N_0}$. Equation \eqref{eq:fkpp} can be derived using a Kramers-Moyal expansion, see \cite{pigolotti2013growth} and chapter 13 in \cite{gardiner1985handbook}. In the derivation, we neglected contributions to the noise coming from the diffusion operator, see \cite{pigolotti2013growth}.

In this case without advection, $\vec{v}=0$, the concentration $c(x,y;t)$ is subject to two competing effects: the noise term in Eq.~\eqref{eq:fkpp} which creates fluctuations around the average solution $\langle c(x,y;t)\rangle=1$, and the diffusion term which smoothens these fluctuations.   

\bibliography{bibliography_sorted_with_authors.bib}

\begin{thebibliography}{43}%
\makeatletter
\providecommand \@ifxundefined [1]{%
 \@ifx{#1\undefined}
}%
\providecommand \@ifnum [1]{%
 \ifnum #1\expandafter \@firstoftwo
 \else \expandafter \@secondoftwo
 \fi
}%
\providecommand \@ifx [1]{%
 \ifx #1\expandafter \@firstoftwo
 \else \expandafter \@secondoftwo
 \fi
}%
\providecommand \natexlab [1]{#1}%
\providecommand \enquote  [1]{``#1''}%
\providecommand \bibnamefont  [1]{#1}%
\providecommand \bibfnamefont [1]{#1}%
\providecommand \citenamefont [1]{#1}%
\providecommand \href@noop [0]{\@secondoftwo}%
\providecommand \href [0]{\begingroup \@sanitize@url \@href}%
\providecommand \@href[1]{\@@startlink{#1}\@@href}%
\providecommand \@@href[1]{\endgroup#1\@@endlink}%
\providecommand \@sanitize@url [0]{\catcode `\\12\catcode `\$12\catcode
  `\&12\catcode `\#12\catcode `\^12\catcode `\_12\catcode `\%12\relax}%
\providecommand \@@startlink[1]{}%
\providecommand \@@endlink[0]{}%
\providecommand \url  [0]{\begingroup\@sanitize@url \@url }%
\providecommand \@url [1]{\endgroup\@href {#1}{\urlprefix }}%
\providecommand \urlprefix  [0]{URL }%
\providecommand \Eprint [0]{\href }%
\providecommand \doibase [0]{https://doi.org/}%
\providecommand \selectlanguage [0]{\@gobble}%
\providecommand \bibinfo  [0]{\@secondoftwo}%
\providecommand \bibfield  [0]{\@secondoftwo}%
\providecommand \translation [1]{[#1]}%
\providecommand \BibitemOpen [0]{}%
\providecommand \bibitemStop [0]{}%
\providecommand \bibitemNoStop [0]{.\EOS\space}%
\providecommand \EOS [0]{\spacefactor3000\relax}%
\providecommand \BibitemShut  [1]{\csname bibitem#1\endcsname}%
\let\auto@bib@innerbib\@empty
%</preamble>
\bibitem [{\citenamefont {Levin}(1992)}]{levin1992problem}%
  \BibitemOpen
  \bibfield  {author} {\bibinfo {author} {\bibfnamefont {S.~A.}\ \bibnamefont
  {Levin}},\ }\bibfield  {title} {\bibinfo {title} {The problem of pattern and
  scale in ecology: the {Robert H. MacArthur award} lecture},\ }\href@noop {}
  {\bibfield  {journal} {\bibinfo  {journal} {Ecology}\ }\textbf {\bibinfo
  {volume} {73}},\ \bibinfo {pages} {1943} (\bibinfo {year}
  {1992})}\BibitemShut {NoStop}%
\bibitem [{\citenamefont {Cencini}\ \emph {et~al.}(2012)\citenamefont
  {Cencini}, \citenamefont {Pigolotti},\ and\ \citenamefont
  {Munoz}}]{cencini2012ecological}%
  \BibitemOpen
  \bibfield  {author} {\bibinfo {author} {\bibfnamefont {M.}~\bibnamefont
  {Cencini}}, \bibinfo {author} {\bibfnamefont {S.}~\bibnamefont {Pigolotti}},\
  and\ \bibinfo {author} {\bibfnamefont {M.~A.}\ \bibnamefont {Munoz}},\
  }\bibfield  {title} {\bibinfo {title} {What ecological factors shape
  species-area curves in neutral models?},\ }\href@noop {} {\bibfield
  {journal} {\bibinfo  {journal} {PloS one}\ }\textbf {\bibinfo {volume} {7}},\
  \bibinfo {pages} {e38232} (\bibinfo {year} {2012})}\BibitemShut {NoStop}%
\bibitem [{\citenamefont {Pigolotti}\ \emph {et~al.}(2018)\citenamefont
  {Pigolotti}, \citenamefont {Cencini}, \citenamefont {Molina},\ and\
  \citenamefont {Mu\~noz}}]{pigolotti2018stochastic}%
  \BibitemOpen
  \bibfield  {author} {\bibinfo {author} {\bibfnamefont {S.}~\bibnamefont
  {Pigolotti}}, \bibinfo {author} {\bibfnamefont {M.}~\bibnamefont {Cencini}},
  \bibinfo {author} {\bibfnamefont {D.}~\bibnamefont {Molina}},\ and\ \bibinfo
  {author} {\bibfnamefont {M.~A.}\ \bibnamefont {Mu\~noz}},\ }\bibfield
  {title} {\bibinfo {title} {Stochastic spatial models in ecology: a
  statistical physics approach},\ }\href@noop {} {\bibfield  {journal}
  {\bibinfo  {journal} {Journal of Statistical Physics}\ }\textbf {\bibinfo
  {volume} {172}},\ \bibinfo {pages} {44} (\bibinfo {year} {2018})}\BibitemShut
  {NoStop}%
\bibitem [{\citenamefont {Durrett}\ and\ \citenamefont
  {Levin}(1996)}]{durrett1996spatial}%
  \BibitemOpen
  \bibfield  {author} {\bibinfo {author} {\bibfnamefont {R.}~\bibnamefont
  {Durrett}}\ and\ \bibinfo {author} {\bibfnamefont {S.}~\bibnamefont
  {Levin}},\ }\bibfield  {title} {\bibinfo {title} {Spatial models for
  species-area curves},\ }\href@noop {} {\bibfield  {journal} {\bibinfo
  {journal} {Journal of Theoretical Biology}\ }\textbf {\bibinfo {volume}
  {179}},\ \bibinfo {pages} {119} (\bibinfo {year} {1996})}\BibitemShut
  {NoStop}%
\bibitem [{\citenamefont {Rosindell}\ and\ \citenamefont
  {Cornell}(2007)}]{rosindell2007species}%
  \BibitemOpen
  \bibfield  {author} {\bibinfo {author} {\bibfnamefont {J.}~\bibnamefont
  {Rosindell}}\ and\ \bibinfo {author} {\bibfnamefont {S.~J.}\ \bibnamefont
  {Cornell}},\ }\bibfield  {title} {\bibinfo {title} {Species--area
  relationships from a spatially explicit neutral model in an infinite
  landscape},\ }\href@noop {} {\bibfield  {journal} {\bibinfo  {journal}
  {Ecology letters}\ }\textbf {\bibinfo {volume} {10}},\ \bibinfo {pages} {586}
  (\bibinfo {year} {2007})}\BibitemShut {NoStop}%
\bibitem [{\citenamefont {Pigolotti}\ and\ \citenamefont
  {Cencini}(2009)}]{pigolotti2009speciation}%
  \BibitemOpen
  \bibfield  {author} {\bibinfo {author} {\bibfnamefont {S.}~\bibnamefont
  {Pigolotti}}\ and\ \bibinfo {author} {\bibfnamefont {M.}~\bibnamefont
  {Cencini}},\ }\bibfield  {title} {\bibinfo {title} {Speciation-rate
  dependence in species--area relationships},\ }\href@noop {} {\bibfield
  {journal} {\bibinfo  {journal} {Journal of theoretical biology}\ }\textbf
  {\bibinfo {volume} {260}},\ \bibinfo {pages} {83} (\bibinfo {year}
  {2009})}\BibitemShut {NoStop}%
\bibitem [{\citenamefont {Durrett}\ and\ \citenamefont
  {Levin}(1994)}]{durrett1994stochastic}%
  \BibitemOpen
  \bibfield  {author} {\bibinfo {author} {\bibfnamefont {R.}~\bibnamefont
  {Durrett}}\ and\ \bibinfo {author} {\bibfnamefont {S.~A.}\ \bibnamefont
  {Levin}},\ }\bibfield  {title} {\bibinfo {title} {Stochastic spatial models:
  a user's guide to ecological applications},\ }\href@noop {} {\bibfield
  {journal} {\bibinfo  {journal} {Philosophical Transactions of the Royal
  Society of London. Series B: Biological Sciences}\ }\textbf {\bibinfo
  {volume} {343}},\ \bibinfo {pages} {329} (\bibinfo {year}
  {1994})}\BibitemShut {NoStop}%
\bibitem [{\citenamefont {Cox}(1989)}]{cox1989coalescing}%
  \BibitemOpen
  \bibfield  {author} {\bibinfo {author} {\bibfnamefont {J.~T.}\ \bibnamefont
  {Cox}},\ }\bibfield  {title} {\bibinfo {title} {Coalescing random walks and
  voter model consensus times on the torus in {$Z^{d}$}},\ }\href@noop {}
  {\bibfield  {journal} {\bibinfo  {journal} {The Annals of Probability}\ ,\
  \bibinfo {pages} {1333}} (\bibinfo {year} {1989})}\BibitemShut {NoStop}%
\bibitem [{\citenamefont {Bramson}\ and\ \citenamefont
  {Lebowitz}(1991)}]{bramson1991asymptotic}%
  \BibitemOpen
  \bibfield  {author} {\bibinfo {author} {\bibfnamefont {M.}~\bibnamefont
  {Bramson}}\ and\ \bibinfo {author} {\bibfnamefont {J.~L.}\ \bibnamefont
  {Lebowitz}},\ }\bibfield  {title} {\bibinfo {title} {Asymptotic behavior of
  densities for two-particle annihilating random walks},\ }\href@noop {}
  {\bibfield  {journal} {\bibinfo  {journal} {Journal of statistical physics}\
  }\textbf {\bibinfo {volume} {62}},\ \bibinfo {pages} {297} (\bibinfo {year}
  {1991})}\BibitemShut {NoStop}%
\bibitem [{\citenamefont {Toroczkai}\ \emph {et~al.}(1998)\citenamefont
  {Toroczkai}, \citenamefont {K{\'a}rolyi}, \citenamefont {P{\'e}ntek},
  \citenamefont {T{\'e}l},\ and\ \citenamefont
  {Grebogi}}]{toroczkai1998advection}%
  \BibitemOpen
  \bibfield  {author} {\bibinfo {author} {\bibfnamefont {Z.}~\bibnamefont
  {Toroczkai}}, \bibinfo {author} {\bibfnamefont {G.}~\bibnamefont
  {K{\'a}rolyi}}, \bibinfo {author} {\bibfnamefont {{\'A}.}~\bibnamefont
  {P{\'e}ntek}}, \bibinfo {author} {\bibfnamefont {T.}~\bibnamefont
  {T{\'e}l}},\ and\ \bibinfo {author} {\bibfnamefont {C.}~\bibnamefont
  {Grebogi}},\ }\bibfield  {title} {\bibinfo {title} {Advection of active
  particles in open chaotic flows},\ }\href@noop {} {\bibfield  {journal}
  {\bibinfo  {journal} {Physical review letters}\ }\textbf {\bibinfo {volume}
  {80}},\ \bibinfo {pages} {500} (\bibinfo {year} {1998})}\BibitemShut
  {NoStop}%
\bibitem [{\citenamefont {K{\'a}rolyi}\ \emph {et~al.}(2000)\citenamefont
  {K{\'a}rolyi}, \citenamefont {P{\'e}ntek}, \citenamefont {Scheuring},
  \citenamefont {T{\'e}l},\ and\ \citenamefont
  {Toroczkai}}]{karolyi2000chaotic}%
  \BibitemOpen
  \bibfield  {author} {\bibinfo {author} {\bibfnamefont {G.}~\bibnamefont
  {K{\'a}rolyi}}, \bibinfo {author} {\bibfnamefont {{\'A}.}~\bibnamefont
  {P{\'e}ntek}}, \bibinfo {author} {\bibfnamefont {I.}~\bibnamefont
  {Scheuring}}, \bibinfo {author} {\bibfnamefont {T.}~\bibnamefont {T{\'e}l}},\
  and\ \bibinfo {author} {\bibfnamefont {Z.}~\bibnamefont {Toroczkai}},\
  }\bibfield  {title} {\bibinfo {title} {Chaotic flow: the physics of species
  coexistence},\ }\href@noop {} {\bibfield  {journal} {\bibinfo  {journal}
  {Proceedings of the National Academy of Sciences}\ }\textbf {\bibinfo
  {volume} {97}},\ \bibinfo {pages} {13661} (\bibinfo {year}
  {2000})}\BibitemShut {NoStop}%
\bibitem [{\citenamefont {Hern{\'a}ndez-Garc{\'\i}a}\ and\ \citenamefont
  {L{\'o}pez}(2004)}]{hernandez2004clustering}%
  \BibitemOpen
  \bibfield  {author} {\bibinfo {author} {\bibfnamefont {E.}~\bibnamefont
  {Hern{\'a}ndez-Garc{\'\i}a}}\ and\ \bibinfo {author} {\bibfnamefont
  {C.}~\bibnamefont {L{\'o}pez}},\ }\bibfield  {title} {\bibinfo {title}
  {Clustering, advection, and patterns in a model of population dynamics with
  neighborhood-dependent rates},\ }\href@noop {} {\bibfield  {journal}
  {\bibinfo  {journal} {Physical Review E}\ }\textbf {\bibinfo {volume} {70}},\
  \bibinfo {pages} {016216} (\bibinfo {year} {2004})}\BibitemShut {NoStop}%
\bibitem [{\citenamefont {Pigolotti}\ \emph {et~al.}(2012)\citenamefont
  {Pigolotti}, \citenamefont {Benzi}, \citenamefont {Jensen},\ and\
  \citenamefont {Nelson}}]{pigolotti2012population}%
  \BibitemOpen
  \bibfield  {author} {\bibinfo {author} {\bibfnamefont {S.}~\bibnamefont
  {Pigolotti}}, \bibinfo {author} {\bibfnamefont {R.}~\bibnamefont {Benzi}},
  \bibinfo {author} {\bibfnamefont {M.~H.}\ \bibnamefont {Jensen}},\ and\
  \bibinfo {author} {\bibfnamefont {D.~R.}\ \bibnamefont {Nelson}},\ }\bibfield
   {title} {\bibinfo {title} {Population genetics in compressible flows},\
  }\href@noop {} {\bibfield  {journal} {\bibinfo  {journal} {Physical review
  letters}\ }\textbf {\bibinfo {volume} {108}},\ \bibinfo {pages} {128102}
  (\bibinfo {year} {2012})}\BibitemShut {NoStop}%
\bibitem [{\citenamefont {Pigolotti}\ \emph {et~al.}(2013)\citenamefont
  {Pigolotti}, \citenamefont {Benzi}, \citenamefont {Perlekar}, \citenamefont
  {Jensen}, \citenamefont {Toschi},\ and\ \citenamefont
  {Nelson}}]{pigolotti2013growth}%
  \BibitemOpen
  \bibfield  {author} {\bibinfo {author} {\bibfnamefont {S.}~\bibnamefont
  {Pigolotti}}, \bibinfo {author} {\bibfnamefont {R.}~\bibnamefont {Benzi}},
  \bibinfo {author} {\bibfnamefont {P.}~\bibnamefont {Perlekar}}, \bibinfo
  {author} {\bibfnamefont {M.~H.}\ \bibnamefont {Jensen}}, \bibinfo {author}
  {\bibfnamefont {F.}~\bibnamefont {Toschi}},\ and\ \bibinfo {author}
  {\bibfnamefont {D.~R.}\ \bibnamefont {Nelson}},\ }\bibfield  {title}
  {\bibinfo {title} {Growth, competition and cooperation in spatial population
  genetics},\ }\href@noop {} {\bibfield  {journal} {\bibinfo  {journal}
  {Theoretical population biology}\ }\textbf {\bibinfo {volume} {84}},\
  \bibinfo {pages} {72} (\bibinfo {year} {2013})}\BibitemShut {NoStop}%
\bibitem [{\citenamefont {Herrer\'ias-Azcu\'e}\ \emph
  {et~al.}(2018)\citenamefont {Herrer\'ias-Azcu\'e}, \citenamefont
  {P\'erez-Mu\~nuzuri},\ and\ \citenamefont {Galla}}]{herrerias2018stirring}%
  \BibitemOpen
  \bibfield  {author} {\bibinfo {author} {\bibfnamefont {F.}~\bibnamefont
  {Herrer\'ias-Azcu\'e}}, \bibinfo {author} {\bibfnamefont {V.}~\bibnamefont
  {P\'erez-Mu\~nuzuri}},\ and\ \bibinfo {author} {\bibfnamefont
  {T.}~\bibnamefont {Galla}},\ }\bibfield  {title} {\bibinfo {title} {Stirring
  does not make populations well mixed},\ }\href@noop {} {\bibfield  {journal}
  {\bibinfo  {journal} {Scientific reports}\ }\textbf {\bibinfo {volume} {8}},\
  \bibinfo {pages} {4068} (\bibinfo {year} {2018})}\BibitemShut {NoStop}%
\bibitem [{\citenamefont {Plummer}\ \emph {et~al.}(2019)\citenamefont
  {Plummer}, \citenamefont {Benzi}, \citenamefont {Nelson},\ and\ \citenamefont
  {Toschi}}]{plummer2019fixation}%
  \BibitemOpen
  \bibfield  {author} {\bibinfo {author} {\bibfnamefont {A.}~\bibnamefont
  {Plummer}}, \bibinfo {author} {\bibfnamefont {R.}~\bibnamefont {Benzi}},
  \bibinfo {author} {\bibfnamefont {D.~R.}\ \bibnamefont {Nelson}},\ and\
  \bibinfo {author} {\bibfnamefont {F.}~\bibnamefont {Toschi}},\ }\bibfield
  {title} {\bibinfo {title} {Fixation probabilities in weakly compressible
  fluid flows},\ }\href@noop {} {\bibfield  {journal} {\bibinfo  {journal}
  {Proceedings of the National Academy of Sciences}\ }\textbf {\bibinfo
  {volume} {116}},\ \bibinfo {pages} {373} (\bibinfo {year}
  {2019})}\BibitemShut {NoStop}%
\bibitem [{\citenamefont {Guccione}\ \emph {et~al.}(2021)\citenamefont
  {Guccione}, \citenamefont {Benzi},\ and\ \citenamefont
  {Toschi}}]{guccione2021strong}%
  \BibitemOpen
  \bibfield  {author} {\bibinfo {author} {\bibfnamefont {G.}~\bibnamefont
  {Guccione}}, \bibinfo {author} {\bibfnamefont {R.}~\bibnamefont {Benzi}},\
  and\ \bibinfo {author} {\bibfnamefont {F.}~\bibnamefont {Toschi}},\
  }\bibfield  {title} {\bibinfo {title} {Strong noise limit for population
  dynamics in incompressible advection},\ }\href@noop {} {\bibfield  {journal}
  {\bibinfo  {journal} {Physical Review E}\ }\textbf {\bibinfo {volume}
  {104}},\ \bibinfo {pages} {034421} (\bibinfo {year} {2021})}\BibitemShut
  {NoStop}%
\bibitem [{\citenamefont {Heinsalu}\ \emph {et~al.}(2013)\citenamefont
  {Heinsalu}, \citenamefont {Hern{\'a}ndez-Garcia},\ and\ \citenamefont
  {L{\'o}pez}}]{heinsalu2013clustering}%
  \BibitemOpen
  \bibfield  {author} {\bibinfo {author} {\bibfnamefont {E.}~\bibnamefont
  {Heinsalu}}, \bibinfo {author} {\bibfnamefont {E.}~\bibnamefont
  {Hern{\'a}ndez-Garcia}},\ and\ \bibinfo {author} {\bibfnamefont
  {C.}~\bibnamefont {L{\'o}pez}},\ }\bibfield  {title} {\bibinfo {title}
  {Clustering determines who survives for competing brownian and l{\'e}vy
  walkers},\ }\href@noop {} {\bibfield  {journal} {\bibinfo  {journal}
  {Physical review letters}\ }\textbf {\bibinfo {volume} {110}},\ \bibinfo
  {pages} {258101} (\bibinfo {year} {2013})}\BibitemShut {NoStop}%
\bibitem [{\citenamefont {Pigolotti}\ and\ \citenamefont
  {Benzi}(2014)}]{pigolotti2014selective}%
  \BibitemOpen
  \bibfield  {author} {\bibinfo {author} {\bibfnamefont {S.}~\bibnamefont
  {Pigolotti}}\ and\ \bibinfo {author} {\bibfnamefont {R.}~\bibnamefont
  {Benzi}},\ }\bibfield  {title} {\bibinfo {title} {Selective advantage of
  diffusing faster},\ }\href@noop {} {\bibfield  {journal} {\bibinfo  {journal}
  {Physical review letters}\ }\textbf {\bibinfo {volume} {112}},\ \bibinfo
  {pages} {188102} (\bibinfo {year} {2014})}\BibitemShut {NoStop}%
\bibitem [{\citenamefont {Pigolotti}\ and\ \citenamefont
  {Benzi}(2016)}]{pigolotti2016competition}%
  \BibitemOpen
  \bibfield  {author} {\bibinfo {author} {\bibfnamefont {S.}~\bibnamefont
  {Pigolotti}}\ and\ \bibinfo {author} {\bibfnamefont {R.}~\bibnamefont
  {Benzi}},\ }\bibfield  {title} {\bibinfo {title} {Competition between
  fast-and slow-diffusing species in non-homogeneous environments},\
  }\href@noop {} {\bibfield  {journal} {\bibinfo  {journal} {Journal of
  theoretical biology}\ }\textbf {\bibinfo {volume} {395}},\ \bibinfo {pages}
  {204} (\bibinfo {year} {2016})}\BibitemShut {NoStop}%
\bibitem [{\citenamefont {Singha}\ \emph {et~al.}(2020)\citenamefont {Singha},
  \citenamefont {Perlekar},\ and\ \citenamefont {Barma}}]{singha2020fixation}%
  \BibitemOpen
  \bibfield  {author} {\bibinfo {author} {\bibfnamefont {T.}~\bibnamefont
  {Singha}}, \bibinfo {author} {\bibfnamefont {P.}~\bibnamefont {Perlekar}},\
  and\ \bibinfo {author} {\bibfnamefont {M.}~\bibnamefont {Barma}},\ }\bibfield
   {title} {\bibinfo {title} {Fixation in competing populations: Diffusion and
  strategies for survival},\ }\href@noop {} {\bibfield  {journal} {\bibinfo
  {journal} {Physical Review Research}\ }\textbf {\bibinfo {volume} {2}},\
  \bibinfo {pages} {023412} (\bibinfo {year} {2020})}\BibitemShut {NoStop}%
\bibitem [{\citenamefont {Bainbridge}(1957)}]{bainbridge1957size}%
  \BibitemOpen
  \bibfield  {author} {\bibinfo {author} {\bibfnamefont {R.}~\bibnamefont
  {Bainbridge}},\ }\bibfield  {title} {\bibinfo {title} {The size, shape and
  density of marine phytoplankton concentrations},\ }\href@noop {} {\bibfield
  {journal} {\bibinfo  {journal} {Biological Reviews}\ }\textbf {\bibinfo
  {volume} {32}},\ \bibinfo {pages} {91} (\bibinfo {year} {1957})}\BibitemShut
  {NoStop}%
\bibitem [{\citenamefont {Scheffer}\ \emph {et~al.}(1995)\citenamefont
  {Scheffer}, \citenamefont {Baveco}, \citenamefont {DeAngelis}, \citenamefont
  {Rose},\ and\ \citenamefont {van Nes}}]{scheffer1995super}%
  \BibitemOpen
  \bibfield  {author} {\bibinfo {author} {\bibfnamefont {M.}~\bibnamefont
  {Scheffer}}, \bibinfo {author} {\bibfnamefont {J.}~\bibnamefont {Baveco}},
  \bibinfo {author} {\bibfnamefont {D.}~\bibnamefont {DeAngelis}}, \bibinfo
  {author} {\bibfnamefont {K.~A.}\ \bibnamefont {Rose}},\ and\ \bibinfo
  {author} {\bibfnamefont {E.}~\bibnamefont {van Nes}},\ }\bibfield  {title}
  {\bibinfo {title} {Super-individuals a simple solution for modelling large
  populations on an individual basis},\ }\href@noop {} {\bibfield  {journal}
  {\bibinfo  {journal} {Ecological modelling}\ }\textbf {\bibinfo {volume}
  {80}},\ \bibinfo {pages} {161} (\bibinfo {year} {1995})}\BibitemShut
  {NoStop}%
\bibitem [{\citenamefont {Villa~Martin}\ \emph {et~al.}(2020)\citenamefont
  {Villa~Martin}, \citenamefont {Bucek}, \citenamefont {Bourguignon},\ and\
  \citenamefont {Pigolotti}}]{villa2020ocean}%
  \BibitemOpen
  \bibfield  {author} {\bibinfo {author} {\bibfnamefont {P.}~\bibnamefont
  {Villa~Martin}}, \bibinfo {author} {\bibfnamefont {A.}~\bibnamefont {Bucek}},
  \bibinfo {author} {\bibfnamefont {T.}~\bibnamefont {Bourguignon}},\ and\
  \bibinfo {author} {\bibfnamefont {S.}~\bibnamefont {Pigolotti}},\ }\bibfield
  {title} {\bibinfo {title} {Ocean currents promote rare species diversity in
  protists},\ }\href@noop {} {\bibfield  {journal} {\bibinfo  {journal}
  {Science Advances}\ }\textbf {\bibinfo {volume} {6}} (\bibinfo {year}
  {2020})}\BibitemShut {NoStop}%
\bibitem [{\citenamefont {Thomas}\ \emph {et~al.}(2008)\citenamefont {Thomas},
  \citenamefont {Tandon},\ and\ \citenamefont
  {Mahadevan}}]{thomas2008submesoscale}%
  \BibitemOpen
  \bibfield  {author} {\bibinfo {author} {\bibfnamefont {L.~N.}\ \bibnamefont
  {Thomas}}, \bibinfo {author} {\bibfnamefont {A.}~\bibnamefont {Tandon}},\
  and\ \bibinfo {author} {\bibfnamefont {A.}~\bibnamefont {Mahadevan}},\
  }\bibfield  {title} {\bibinfo {title} {Submesoscale processes and dynamics},\
  }\href@noop {} {\bibfield  {journal} {\bibinfo  {journal} {Ocean modeling in
  an Eddying Regime}\ }\textbf {\bibinfo {volume} {177}},\ \bibinfo {pages}
  {17} (\bibinfo {year} {2008})}\BibitemShut {NoStop}%
\bibitem [{\citenamefont {Benzi}\ \emph {et~al.}(2012)\citenamefont {Benzi},
  \citenamefont {Jensen}, \citenamefont {Nelson}, \citenamefont {Perlekar},
  \citenamefont {Pigolotti},\ and\ \citenamefont
  {Toschi}}]{benzi2012population}%
  \BibitemOpen
  \bibfield  {author} {\bibinfo {author} {\bibfnamefont {R.}~\bibnamefont
  {Benzi}}, \bibinfo {author} {\bibfnamefont {M.~H.}\ \bibnamefont {Jensen}},
  \bibinfo {author} {\bibfnamefont {D.~R.}\ \bibnamefont {Nelson}}, \bibinfo
  {author} {\bibfnamefont {P.}~\bibnamefont {Perlekar}}, \bibinfo {author}
  {\bibfnamefont {S.}~\bibnamefont {Pigolotti}},\ and\ \bibinfo {author}
  {\bibfnamefont {F.}~\bibnamefont {Toschi}},\ }\bibfield  {title} {\bibinfo
  {title} {Population dynamics in compressible flows},\ }\href@noop {}
  {\bibfield  {journal} {\bibinfo  {journal} {The European Physical Journal
  Special Topics}\ }\textbf {\bibinfo {volume} {204}},\ \bibinfo {pages} {57}
  (\bibinfo {year} {2012})}\BibitemShut {NoStop}%
\bibitem [{\citenamefont {Plummer}\ \emph {et~al.}(2022)\citenamefont
  {Plummer}, \citenamefont {Freilich}, \citenamefont {Benzi}, \citenamefont
  {Choi}, \citenamefont {Sudek}, \citenamefont {Worden}, \citenamefont
  {Toschi},\ and\ \citenamefont {Mahadevan}}]{plummer2022oceanic}%
  \BibitemOpen
  \bibfield  {author} {\bibinfo {author} {\bibfnamefont {A.}~\bibnamefont
  {Plummer}}, \bibinfo {author} {\bibfnamefont {M.}~\bibnamefont {Freilich}},
  \bibinfo {author} {\bibfnamefont {R.}~\bibnamefont {Benzi}}, \bibinfo
  {author} {\bibfnamefont {C.~J.}\ \bibnamefont {Choi}}, \bibinfo {author}
  {\bibfnamefont {L.}~\bibnamefont {Sudek}}, \bibinfo {author} {\bibfnamefont
  {A.~Z.}\ \bibnamefont {Worden}}, \bibinfo {author} {\bibfnamefont
  {F.}~\bibnamefont {Toschi}},\ and\ \bibinfo {author} {\bibfnamefont
  {A.}~\bibnamefont {Mahadevan}},\ }\bibfield  {title} {\bibinfo {title}
  {Oceanic frontal divergence alters phytoplankton competition and
  distribution},\ }\href@noop {} {\bibfield  {journal} {\bibinfo  {journal}
  {arXiv preprint arXiv:2202.11745}\ } (\bibinfo {year} {2022})}\BibitemShut
  {NoStop}%
\bibitem [{\citenamefont {Rosindell}\ \emph {et~al.}(2008)\citenamefont
  {Rosindell}, \citenamefont {Wong},\ and\ \citenamefont
  {Etienne}}]{rosindell2008coalescence}%
  \BibitemOpen
  \bibfield  {author} {\bibinfo {author} {\bibfnamefont {J.}~\bibnamefont
  {Rosindell}}, \bibinfo {author} {\bibfnamefont {Y.}~\bibnamefont {Wong}},\
  and\ \bibinfo {author} {\bibfnamefont {R.~S.}\ \bibnamefont {Etienne}},\
  }\bibfield  {title} {\bibinfo {title} {A coalescence approach to spatial
  neutral ecology},\ }\href@noop {} {\bibfield  {journal} {\bibinfo  {journal}
  {Ecological Informatics}\ }\textbf {\bibinfo {volume} {3}},\ \bibinfo {pages}
  {259} (\bibinfo {year} {2008})}\BibitemShut {NoStop}%
\bibitem [{\citenamefont {Xu}\ \emph {et~al.}(2020)\citenamefont {Xu},
  \citenamefont {B{\"o}ttcher},\ and\ \citenamefont {Chou}}]{xu2020diversity}%
  \BibitemOpen
  \bibfield  {author} {\bibinfo {author} {\bibfnamefont {S.}~\bibnamefont
  {Xu}}, \bibinfo {author} {\bibfnamefont {L.}~\bibnamefont {B{\"o}ttcher}},\
  and\ \bibinfo {author} {\bibfnamefont {T.}~\bibnamefont {Chou}},\ }\bibfield
  {title} {\bibinfo {title} {Diversity in biology: definitions, quantification
  and models},\ }\href@noop {} {\bibfield  {journal} {\bibinfo  {journal}
  {Physical Biology}\ }\textbf {\bibinfo {volume} {17}},\ \bibinfo {pages}
  {031001} (\bibinfo {year} {2020})}\BibitemShut {NoStop}%
\bibitem [{\citenamefont {Whittaker}(1960)}]{whittaker1960vegetation}%
  \BibitemOpen
  \bibfield  {author} {\bibinfo {author} {\bibfnamefont {R.~H.}\ \bibnamefont
  {Whittaker}},\ }\bibfield  {title} {\bibinfo {title} {Vegetation of the
  siskiyou mountains, oregon and california},\ }\href@noop {} {\bibfield
  {journal} {\bibinfo  {journal} {Ecological monographs}\ }\textbf {\bibinfo
  {volume} {30}},\ \bibinfo {pages} {279} (\bibinfo {year} {1960})}\BibitemShut
  {NoStop}%
\bibitem [{\citenamefont
  {Tuomisto}(2010{\natexlab{a}})}]{tuomisto2010diversity1}%
  \BibitemOpen
  \bibfield  {author} {\bibinfo {author} {\bibfnamefont {H.}~\bibnamefont
  {Tuomisto}},\ }\bibfield  {title} {\bibinfo {title} {A diversity of beta
  diversities: straightening up a concept gone awry. {Part 1}. {Defining} beta
  diversity as a function of alpha and gamma diversity},\ }\href@noop {}
  {\bibfield  {journal} {\bibinfo  {journal} {Ecography}\ }\textbf {\bibinfo
  {volume} {33}},\ \bibinfo {pages} {2} (\bibinfo {year}
  {2010}{\natexlab{a}})}\BibitemShut {NoStop}%
\bibitem [{\citenamefont
  {Tuomisto}(2010{\natexlab{b}})}]{tuomisto2010diversity2}%
  \BibitemOpen
  \bibfield  {author} {\bibinfo {author} {\bibfnamefont {H.}~\bibnamefont
  {Tuomisto}},\ }\bibfield  {title} {\bibinfo {title} {A diversity of beta
  diversities: straightening up a concept gone awry. {Part 2}. {Quantifying}
  beta diversity and related phenomena},\ }\href@noop {} {\bibfield  {journal}
  {\bibinfo  {journal} {Ecography}\ }\textbf {\bibinfo {volume} {33}},\
  \bibinfo {pages} {23} (\bibinfo {year} {2010}{\natexlab{b}})}\BibitemShut
  {NoStop}%
\bibitem [{\citenamefont {Rosenzweig}\ \emph {et~al.}(1995)\citenamefont
  {Rosenzweig} \emph {et~al.}}]{rosenzweig1995species}%
  \BibitemOpen
  \bibfield  {author} {\bibinfo {author} {\bibfnamefont {M.~L.}\ \bibnamefont
  {Rosenzweig}} \emph {et~al.},\ }\href@noop {} {\emph {\bibinfo {title}
  {Species diversity in space and time}}}\ (\bibinfo  {publisher} {Cambridge
  University Press},\ \bibinfo {year} {1995})\BibitemShut {NoStop}%
\bibitem [{\citenamefont {Preston}(1960)}]{preston1960time}%
  \BibitemOpen
  \bibfield  {author} {\bibinfo {author} {\bibfnamefont {F.}~\bibnamefont
  {Preston}},\ }\bibfield  {title} {\bibinfo {title} {Time and space and the
  variation of species},\ }\href@noop {} {\bibfield  {journal} {\bibinfo
  {journal} {Ecology}\ }\textbf {\bibinfo {volume} {41}},\ \bibinfo {pages}
  {612} (\bibinfo {year} {1960})}\BibitemShut {NoStop}%
\bibitem [{\citenamefont {Hubbell}(2001)}]{hubbell2001unified}%
  \BibitemOpen
  \bibfield  {author} {\bibinfo {author} {\bibfnamefont {S.~P.}\ \bibnamefont
  {Hubbell}},\ }\href@noop {} {\emph {\bibinfo {title} {The unified neutral
  theory of biodiversity and biogeography (MPB-32)}}}\ (\bibinfo  {publisher}
  {Princeton University Press},\ \bibinfo {year} {2001})\BibitemShut {NoStop}%
\bibitem [{\citenamefont {Arrhenius}(1921)}]{arrhenius1921species}%
  \BibitemOpen
  \bibfield  {author} {\bibinfo {author} {\bibfnamefont {O.}~\bibnamefont
  {Arrhenius}},\ }\bibfield  {title} {\bibinfo {title} {Species and area},\
  }\href@noop {} {\bibfield  {journal} {\bibinfo  {journal} {Journal of
  Ecology}\ }\textbf {\bibinfo {volume} {9}},\ \bibinfo {pages} {95} (\bibinfo
  {year} {1921})}\BibitemShut {NoStop}%
\bibitem [{\citenamefont {Volkov}\ \emph {et~al.}(2003)\citenamefont {Volkov},
  \citenamefont {Banavar}, \citenamefont {Hubbell},\ and\ \citenamefont
  {Maritan}}]{volkov2003neutral}%
  \BibitemOpen
  \bibfield  {author} {\bibinfo {author} {\bibfnamefont {I.}~\bibnamefont
  {Volkov}}, \bibinfo {author} {\bibfnamefont {J.~R.}\ \bibnamefont {Banavar}},
  \bibinfo {author} {\bibfnamefont {S.~P.}\ \bibnamefont {Hubbell}},\ and\
  \bibinfo {author} {\bibfnamefont {A.}~\bibnamefont {Maritan}},\ }\bibfield
  {title} {\bibinfo {title} {Neutral theory and relative species abundance in
  ecology},\ }\href@noop {} {\bibfield  {journal} {\bibinfo  {journal}
  {Nature}\ }\textbf {\bibinfo {volume} {424}},\ \bibinfo {pages} {1035}
  (\bibinfo {year} {2003})}\BibitemShut {NoStop}%
\bibitem [{\citenamefont {Cencini}\ \emph {et~al.}(1999)\citenamefont
  {Cencini}, \citenamefont {Lacorata}, \citenamefont {Vulpiani},\ and\
  \citenamefont {Zambianchi}}]{cencini1999mixing}%
  \BibitemOpen
  \bibfield  {author} {\bibinfo {author} {\bibfnamefont {M.}~\bibnamefont
  {Cencini}}, \bibinfo {author} {\bibfnamefont {G.}~\bibnamefont {Lacorata}},
  \bibinfo {author} {\bibfnamefont {A.}~\bibnamefont {Vulpiani}},\ and\
  \bibinfo {author} {\bibfnamefont {E.}~\bibnamefont {Zambianchi}},\ }\bibfield
   {title} {\bibinfo {title} {Mixing in a meandering jet: A markovian
  approximation},\ }\href@noop {} {\bibfield  {journal} {\bibinfo  {journal}
  {Journal of physical oceanography}\ }\textbf {\bibinfo {volume} {29}},\
  \bibinfo {pages} {2578} (\bibinfo {year} {1999})}\BibitemShut {NoStop}%
\bibitem [{\citenamefont {Biferale}\ \emph {et~al.}(1995)\citenamefont
  {Biferale}, \citenamefont {Crisanti}, \citenamefont {Vergassola},\ and\
  \citenamefont {Vulpiani}}]{biferale1995eddy}%
  \BibitemOpen
  \bibfield  {author} {\bibinfo {author} {\bibfnamefont {L.}~\bibnamefont
  {Biferale}}, \bibinfo {author} {\bibfnamefont {A.}~\bibnamefont {Crisanti}},
  \bibinfo {author} {\bibfnamefont {M.}~\bibnamefont {Vergassola}},\ and\
  \bibinfo {author} {\bibfnamefont {A.}~\bibnamefont {Vulpiani}},\ }\bibfield
  {title} {\bibinfo {title} {Eddy diffusivities in scalar transport},\
  }\href@noop {} {\bibfield  {journal} {\bibinfo  {journal} {Physics of
  Fluids}\ }\textbf {\bibinfo {volume} {7}},\ \bibinfo {pages} {2725} (\bibinfo
  {year} {1995})}\BibitemShut {NoStop}%
\bibitem [{\citenamefont {Ser-Giacomi}\ \emph {et~al.}(2018)\citenamefont
  {Ser-Giacomi}, \citenamefont {Zinger}, \citenamefont {Malviya}, \citenamefont
  {De~Vargas}, \citenamefont {Karsenti}, \citenamefont {Bowler},\ and\
  \citenamefont {De~Monte}}]{ser2018ubiquitous}%
  \BibitemOpen
  \bibfield  {author} {\bibinfo {author} {\bibfnamefont {E.}~\bibnamefont
  {Ser-Giacomi}}, \bibinfo {author} {\bibfnamefont {L.}~\bibnamefont {Zinger}},
  \bibinfo {author} {\bibfnamefont {S.}~\bibnamefont {Malviya}}, \bibinfo
  {author} {\bibfnamefont {C.}~\bibnamefont {De~Vargas}}, \bibinfo {author}
  {\bibfnamefont {E.}~\bibnamefont {Karsenti}}, \bibinfo {author}
  {\bibfnamefont {C.}~\bibnamefont {Bowler}},\ and\ \bibinfo {author}
  {\bibfnamefont {S.}~\bibnamefont {De~Monte}},\ }\bibfield  {title} {\bibinfo
  {title} {Ubiquitous abundance distribution of non-dominant plankton across
  the global ocean},\ }\href@noop {} {\bibfield  {journal} {\bibinfo  {journal}
  {Nature ecology {\&} evolution}\ ,\ \bibinfo {pages} {1}} (\bibinfo {year}
  {2018})}\BibitemShut {NoStop}%
\bibitem [{\citenamefont {Boenigk}\ \emph {et~al.}(2018)\citenamefont
  {Boenigk}, \citenamefont {Wodniok}, \citenamefont {Bock}, \citenamefont
  {Beisser}, \citenamefont {Hempel}, \citenamefont {Grossmann}, \citenamefont
  {Lange},\ and\ \citenamefont {Jensen}}]{boenigk2018geographic}%
  \BibitemOpen
  \bibfield  {author} {\bibinfo {author} {\bibfnamefont {J.}~\bibnamefont
  {Boenigk}}, \bibinfo {author} {\bibfnamefont {S.}~\bibnamefont {Wodniok}},
  \bibinfo {author} {\bibfnamefont {C.}~\bibnamefont {Bock}}, \bibinfo {author}
  {\bibfnamefont {D.}~\bibnamefont {Beisser}}, \bibinfo {author} {\bibfnamefont
  {C.}~\bibnamefont {Hempel}}, \bibinfo {author} {\bibfnamefont
  {L.}~\bibnamefont {Grossmann}}, \bibinfo {author} {\bibfnamefont
  {A.}~\bibnamefont {Lange}},\ and\ \bibinfo {author} {\bibfnamefont
  {M.}~\bibnamefont {Jensen}},\ }\bibfield  {title} {\bibinfo {title}
  {Geographic distance and mountain ranges structure freshwater protist
  communities on a european scale},\ }\href@noop {} {\bibfield  {journal}
  {\bibinfo  {journal} {Metabarcoding and Metagenomics}\ }\textbf {\bibinfo
  {volume} {2}},\ \bibinfo {pages} {e21519} (\bibinfo {year}
  {2018})}\BibitemShut {NoStop}%
\bibitem [{\citenamefont {De~Vargas}\ \emph {et~al.}(2015)\citenamefont
  {De~Vargas}, \citenamefont {Audic}, \citenamefont {Henry}, \citenamefont
  {Decelle}, \citenamefont {Mah\'e}, \citenamefont {Logares}, \citenamefont
  {Lara}, \citenamefont {Berney}, \citenamefont {Le~Bescot}, \citenamefont
  {Probert} \emph {et~al.}}]{de2015eukaryotic}%
  \BibitemOpen
  \bibfield  {author} {\bibinfo {author} {\bibfnamefont {C.}~\bibnamefont
  {De~Vargas}}, \bibinfo {author} {\bibfnamefont {S.}~\bibnamefont {Audic}},
  \bibinfo {author} {\bibfnamefont {N.}~\bibnamefont {Henry}}, \bibinfo
  {author} {\bibfnamefont {J.}~\bibnamefont {Decelle}}, \bibinfo {author}
  {\bibfnamefont {F.}~\bibnamefont {Mah\'e}}, \bibinfo {author} {\bibfnamefont
  {R.}~\bibnamefont {Logares}}, \bibinfo {author} {\bibfnamefont
  {E.}~\bibnamefont {Lara}}, \bibinfo {author} {\bibfnamefont {C.}~\bibnamefont
  {Berney}}, \bibinfo {author} {\bibfnamefont {N.}~\bibnamefont {Le~Bescot}},
  \bibinfo {author} {\bibfnamefont {I.}~\bibnamefont {Probert}}, \emph
  {et~al.},\ }\bibfield  {title} {\bibinfo {title} {Eukaryotic plankton
  diversity in the sunlit ocean},\ }\href@noop {} {\bibfield  {journal}
  {\bibinfo  {journal} {Science}\ }\textbf {\bibinfo {volume} {348}},\ \bibinfo
  {pages} {1261605} (\bibinfo {year} {2015})}\BibitemShut {NoStop}%
\bibitem [{\citenamefont {Gardiner}\ \emph {et~al.}(1985)\citenamefont
  {Gardiner} \emph {et~al.}}]{gardiner1985handbook}%
  \BibitemOpen
  \bibfield  {author} {\bibinfo {author} {\bibfnamefont {C.~W.}\ \bibnamefont
  {Gardiner}} \emph {et~al.},\ }\href@noop {} {\emph {\bibinfo {title}
  {Handbook of stochastic methods}}},\ Vol.~\bibinfo {volume} {3}\ (\bibinfo
  {publisher} {springer Berlin},\ \bibinfo {year} {1985})\BibitemShut {NoStop}%
\end{thebibliography}%



\end{document}